\pdfoutput=1
\documentclass[preprint,preprintnumbers,amsmath,amssymb,floatfix,endfloats*]{revtex4}

\usepackage{graphicx}
\usepackage{dcolumn}
\usepackage{bm}
\usepackage{amsfonts}

\DeclareMathAlphabet{\mathsfsl}{OT1}{cmr}{bx}{it}
\begin{document}
\title{A delayed yielding transition in mechanically annealed binary glasses at finite temperature}
\author{Nikolai V. Priezjev$^{1,2}$}
\affiliation{$^{1}$Department of Mechanical and Materials
Engineering, Wright State University, Dayton, OH 45435}
\affiliation{$^{2}$National Research University Higher School of
Economics, Moscow 101000, Russia}
\date{\today}
\begin{abstract}

The influence of strain amplitude, glass stability and thermal
fluctuations on shear band formation and yielding transition is
studied using molecular dynamics simulations. The model binary
mixture is first gradually cooled below the glass transition
temperature and then periodically deformed to access a broad range
of potential energy states. We find that the critical strain
amplitude becomes larger for highly annealed glasses within about
one thousand shear cycles. Moreover, upon continued loading at a
fixed strain amplitude, the yielding transition is delayed in
glasses mechanically annealed to lower energy states.  It is also
demonstrated that nucleation of a small cluster of atoms with large
nonaffine displacements precedes a sharp energy change associated
with the yielding transition. These results are important for
thermal and mechanical processing of amorphous alloys with tunable
mechanical and physical properties.

\vskip 0.5in


\end{abstract}

\maketitle

\section{Introduction}

Understanding how the disordered atomic structure of amorphous
alloys evolves under thermal or mechanical treatments is important
for predicting their physical and mechanical
properties~\cite{Qiao19}.  For instance, due to their enhanced
strength, elastic limit and corrosion resistance, bulk metallic
glasses become promising materials for medical and dental
applications~\cite{Imai20}.  It is well understood by now that
plastic deformation in disordered solids proceeds via collective
rearrangements of small groups of atoms or shear
transformations~\cite{Spaepen77,Argon79}. Depending on the degree of
annealing, metallic glasses can become brittle and fail via sudden
formation of narrow regions where strain is localized within the
so-called shear bands.  In turn, thermo-mechanical processing might
lead to higher energy, rejuvenated states and, thus, improved
plasticity~\cite{Greer16}. The common techniques include high
pressure torsion, elastostatic loading, shot peening, ion
irradiation, and more recently discovered thermal
cycling~\cite{Greer16,Ketov15}.   However, design and development of
novel processing routes for amorphous alloys with optimized
properties are impeded by the time required to explore the vast
parameter space.

\vskip 0.05in

In recent years, the structural relaxation dynamics and yielding
transition in amorphous materials under periodic loading conditions
were extensively studied using atomistic
simulations~\cite{Priezjev13,Sastry13,Reichhardt13,Priezjev14,
IdoNature15, Priezjev16, Kawasaki16, Priezjev16a, Sastry17,
Priezjev17, OHern17, Priezjev18, Priezjev18a, NVP18strload,
Sastry18, PriMakrho05, PriMakrho09, Sastry19band, PriezSHALT19,
Ido2020, Priez19ba, Peng20, Jana20, KawBer19, BhaSastry19,
Priez20alt}. Remarkably, it was found that following a number of
transient deformation cycles, the particle trajectories in a
disordered solid at zero temperature become exactly reversible and
fall into the so-called `limit cycles'~\cite{Reichhardt13,
IdoNature15}. In general, the potential energy of thermal glasses
can be reduced during small-amplitude periodic loading via a series
of collective, irreversible rearrangements of atoms, in a process
termed `mechanical annealing'~\cite{Priezjev18, NVP18strload,
Sastry18, PriMakrho05, PriMakrho09, Sastry19band, PriezSHALT19,
Jana20}.  By contrast, the oscillatory deformation at sufficiently
large strain amplitudes leads to the formation of shear bands, while
the onset of yielding is accelerated at higher temperatures, larger
strain amplitudes, or due to alternating shear
orientation~\cite{Sastry13, Kawasaki16, Sastry17, Priezjev17,
Priezjev18a, PriMakrho09, Sastry19band, Priez19ba, Priez20alt}. More
recently, it was demonstrated that the critical strain amplitude at
zero temperature becomes larger in highly stable glasses obtained
using either the swap Monte Carlo algorithm~\cite{KawBer19} or
mechanical annealing~\cite{BhaSastry19}. In spite of these efforts,
however, the interplay between glass stability, thermal
fluctuations, and mechanical deformation in relation to strain
localization and yielding remains largely unexplored.

\vskip 0.05in

In this paper, the effects of temperature, degree of annealing, and
strain amplitude on the yielding transition in amorphous alloys are
investigated using molecular dynamics simulations. The model glass
is represented by the binary mixture, which is gradually cooled
below the glass transition temperature and then further annealed via
small-amplitude cyclic loading to progressively lower energy states.
It will be shown that the critical strain amplitude increases for
more stable glasses loaded during about one thousand cycles at
finite temperatures.  On the other hand, the yielding transition is
delayed in lower energy glasses subjected to thousands of shear
cycles with a fixed strain amplitude.

\vskip 0.05in

The contents of this paper are as follows. We describe the details
of molecular dynamics simulations and the deformation protocol in
the next section. The numerical results for the mechanical
annealing, potential energy, mechanical properties, and shear band
formation during the yielding transition in periodically driven
binary glasses are presented in section\,\ref{sec:Results}.  The
brief summary is given in the last section.

\section{Molecular dynamics (MD) simulations}
\label{sec:MD_Model}

The composition of the amorphous alloy consists of (80:20) binary
mixture with strongly non-additive interaction between different
types of atoms, which suppresses crystallization upon
cooling~\cite{KobAnd95}. The model was introduced by Kob and
Andersen (KA) about twenty years ago and has since been extensively
studied by a number of groups~\cite{KobAnd95}. In the KA model, the
interaction between any two atoms is specified via the Lennard-Jones
(LJ) potential:
\begin{equation}
V_{\alpha\beta}(r)=4\,\varepsilon_{\alpha\beta}\,\Big[\Big(\frac{\sigma_{\alpha\beta}}{r}\Big)^{12}\!-
\Big(\frac{\sigma_{\alpha\beta}}{r}\Big)^{6}\,\Big],
\label{Eq:LJ_KA}
\end{equation}
with the parameters: $\varepsilon_{AA}=1.0$, $\varepsilon_{AB}=1.5$,
$\varepsilon_{BB}=0.5$, $\sigma_{AA}=1.0$, $\sigma_{AB}=0.8$,
$\sigma_{BB}=0.88$, and $m_{A}=m_{B}$~\cite{KobAnd95}. We note that
this parametrization is similar to the description of the amorphous
metal-metalloid alloy $\text{Ni}_{80}\text{P}_{20}$ studied by Weber
and Stillinger~\cite{Weber85}.   In the present study, the LJ
potential was truncated at the cutoff radius
$r_{c,\,\alpha\beta}=2.5\,\sigma_{\alpha\beta}$ to reduce
computational efforts.  The simulation domain contains $48\,000$
type $A$ atoms and $12\,000$ type $B$ atoms, and the total number of
atoms $60\,000$ is kept fixed.   The following reduced units of
length, mass, and energy $\sigma=\sigma_{AA}$, $m=m_{A}$, and
$\varepsilon=\varepsilon_{AA}$ were used for all physical
quantities.   Furthermore, the velocity Verlet algorithm with the
time step $\triangle t_{MD}=0.005\,\tau$ was employed for the
numerical integration of the equations of
motion~\cite{Allen87,Lammps}. Here,
$\tau=\sigma\sqrt{m/\varepsilon}$ is the characteristic LJ time.

\vskip 0.05in


The binary mixture was first thoroughly equilibrated at the
temperature $T_{LJ}=1.0\,\varepsilon/k_B$ and density
$\rho=\rho_A+\rho_B=1.2\,\sigma^{-3}$. Here, $k_B$ denotes the
Boltzmann constant. For reference, the critical temperature of the
KA model at this density is
$T_c=0.435\,\varepsilon/k_B$~\cite{KobAnd95}. The temperature was
regulated via the Nos\'{e}-Hoover thermostat~\cite{Allen87,Lammps}.
All simulations were carried out at the constant volume, and the
size of the periodic box was fixed to $L=36.84\,\sigma$. Following
the equilibration period, the system was linearly cooled below the
glass transition at the rate $10^{-5}\varepsilon/k_{B}\tau$ to
$T_{LJ}=0.30\,\varepsilon/k_B$.

\vskip 0.05in


In order to access lower potential energy states, periodic shear
deformation along the $xz$ plane was imposed as follows:
\begin{equation}
\gamma(t)=\gamma_0\,\text{sin}(2\pi t/T),
\label{Eq:shear}
\end{equation}
where $\gamma_0$ is the strain amplitude and $T=5000\,\tau$ is the
oscillation period. The cyclic loading was first applied at
$T_{LJ}=0.30\,\varepsilon/k_B$ and the strain amplitude
$\gamma_0=0.035$.   It was previously shown that this value of
strain amplitude is below the yielding strain at
$T_{LJ}=0.30\,\varepsilon/k_B$ and $\rho=1.2\,\sigma^{-3}$, and the
relaxation rate (the potential energy decrease over time) is
relatively large~\cite{Sastry18}. Further, the dynamic response of
the annealed glass to oscillatory shear deformation was investigated
at two temperatures, namely, $T_{LJ}=0.1\,\varepsilon/k_B$ and
$0.01\,\varepsilon/k_B$. Due to computational limitations, the MD
simulations were performed only for one sample. The typical
production run during 3400 cycles with the period $T=5000\,\tau$
required about 94 days using 40 processors in parallel.

\section{Results}
\label{sec:Results}


Amorphous alloys like metallic glasses are typically formed upon
rapid cooling across the glass transition temperature, and, due to
the absence of topological defects, the yield stress during
deformation is relatively large~\cite{Ma11}.  The stress overshoot
can be further increased by relocating the glass to lower potential
energy states via thermal or mechanical annealing~\cite{Greer16}. In
particular, it was recently shown that the critical strain amplitude
increases in well annealed glasses subjected to athermal quasistaic
cyclic shear deformation~\cite{KawBer19,BhaSastry19}. While in
athermal simulations particle trajectories become exactly reversible
(after a certain number of cycles) at strain amplitudes below the
yielding transition, the periodic deformation at finite temperatures
might instead lead to a delay in yielding.   In other words, the
number of cycles required to form a shear band in the presence of
thermal fluctuations might depend on the degree of annealing as well
as the strain amplitude. In what follows, the dynamic behavior of
periodically deformed binary glasses is studied at strain amplitudes
in the vicinity of the yielding transition at finite temperatures
and in a wide range of potential energy states.

\vskip 0.05in

\begin{figure}[t]
\includegraphics[width=12.0cm,angle=0]{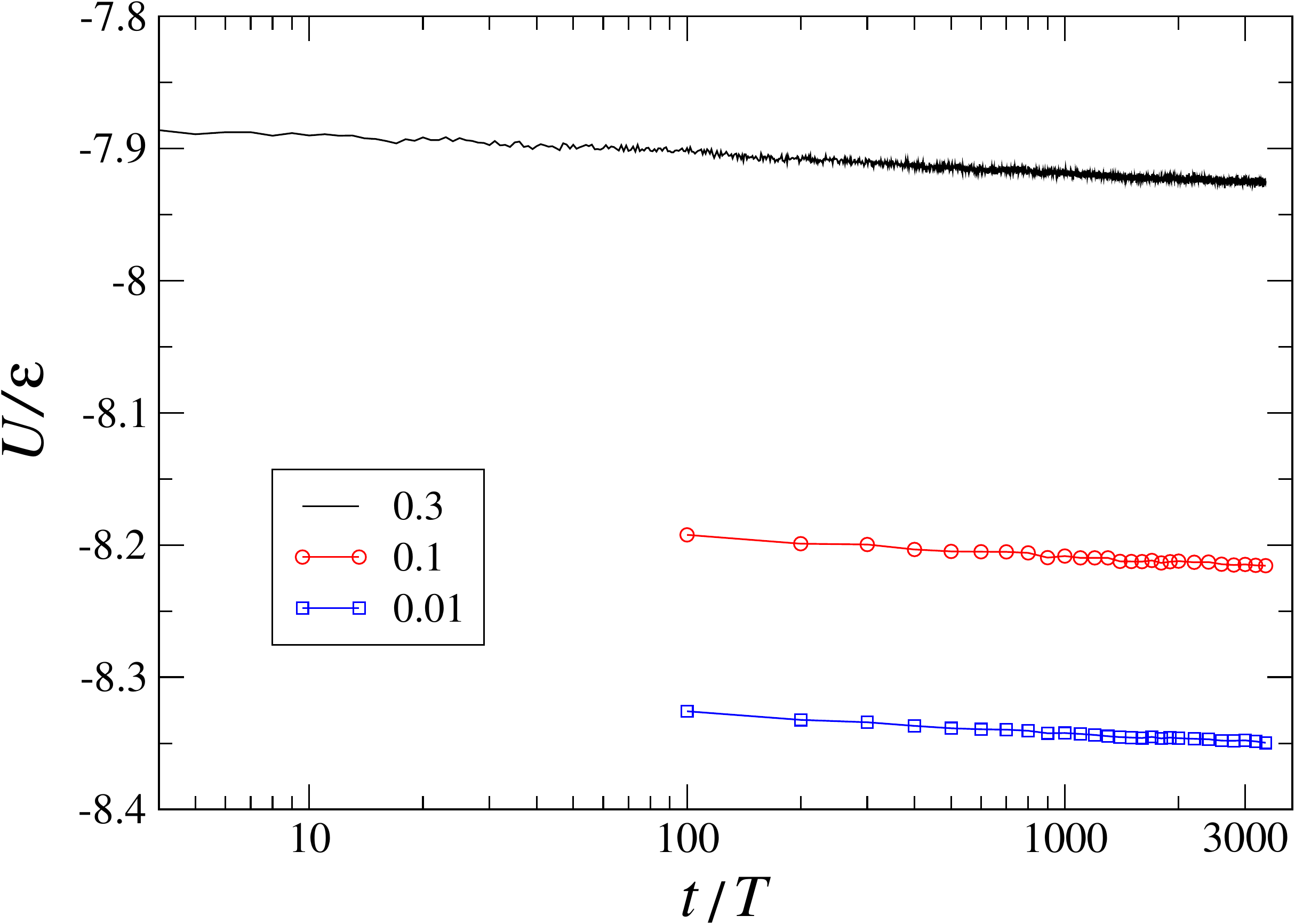}
\caption{(Color online) The potential energy minima (at zero strain)
during cyclic shear with the strain amplitude $\gamma_0=0.035$ at
the temperature $T_{LJ}=0.3\,\varepsilon/k_B$ (black line). The
potential energy of binary glasses (at zero strain) annealed to the
temperatures $T_{LJ}=0.1\,\varepsilon/k_B$ (red circles) and
$T_{LJ}=0.01\,\varepsilon/k_B$ (blue squares). The period of
oscillation is $T=5000\,\tau$.}
\label{fig:poten_mech_ann}
\end{figure}

The potential energy per atom at the end of each cycle is reported
in Fig.\,\ref{fig:poten_mech_ann} for the strain amplitude
$\gamma_0=0.035$ and temperature $T_{LJ}=0.30\,\varepsilon/k_B$.
Note that the black line in Fig.\,\ref{fig:poten_mech_ann} denotes
the energy minima when the imposed strain is zero. It is clearly
seen that the potential energy continues to decay monotonically with
increasing cycle number, and the plateau level is yet to be reached.
The energy decrease during 3400 cycles at
$T_{LJ}=0.30\,\varepsilon/k_B$ is $\Delta U\approx
0.046\,\varepsilon$.  Next, after a certain number of cycles when
strain is zero, the binary glass was cooled from
$T_{LJ}=0.30\,\varepsilon/k_B$ to either
$T_{LJ}=0.10\,\varepsilon/k_B$ or $0.01\,\varepsilon/k_B$ during the
time interval $10^4\,\tau$. The resulting energy levels at the
selected cycles are shown in Fig.\,\ref{fig:poten_mech_ann} by red
circles and blue squares, respectively.  As is evident, the slopes
of the energy decay are nearly the same at all temperatures
indicating that the inherent structure remains essentially unchanged
upon cooling to lower temperatures.  In the following, the yielding
transition and strain localization during oscillatory shear
deformation will be examined at two temperatures,
$T_{LJ}=0.10\,\varepsilon/k_B$ and $0.01\,\varepsilon/k_B$, well
below the glass transition temperature
$\approx\!0.435\,\varepsilon/k_B$.

\vskip 0.05in

\begin{figure}[t]
\includegraphics[width=12.0cm,angle=0]{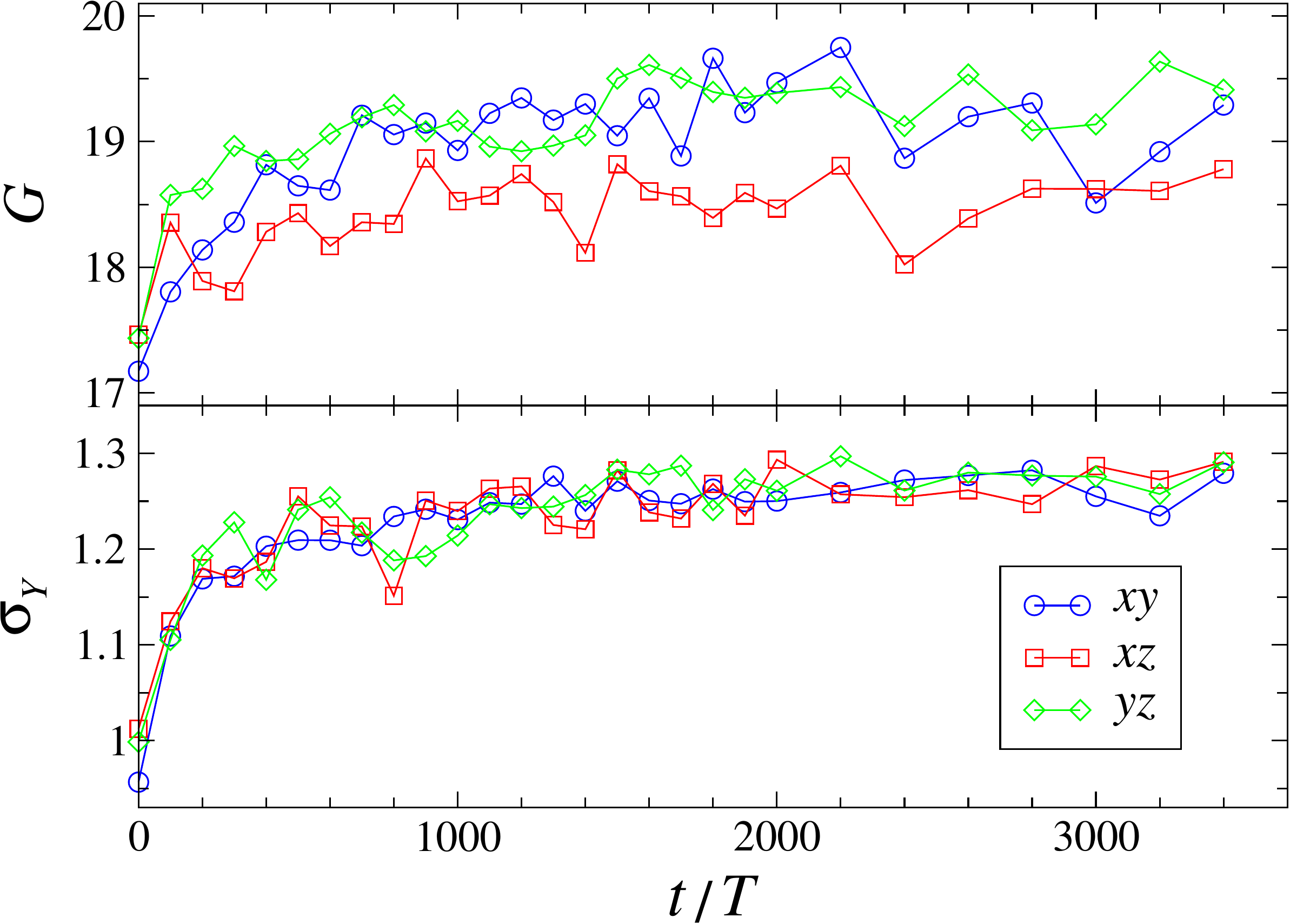}
\caption{(Color online)  The shear modulus $G$ (in units of
$\varepsilon\sigma^{-3}$) and the yielding peak $\sigma_Y$ (in units
of $\varepsilon\sigma^{-3}$) versus cycle number at the temperature
$T_{LJ}=0.1\,\varepsilon/k_B$. The startup continuous shear
deformation with the strain rate $\dot{\gamma}=10^{-5}\,\tau^{-1}$
is applied along the $xy$ plane (blue circles), $xz$ plane (red
squares), and $yz$ plane (green diamonds). }
\label{fig:G_and_Y_T01}
\end{figure}
\begin{figure}[t]
\includegraphics[width=12.0cm,angle=0]{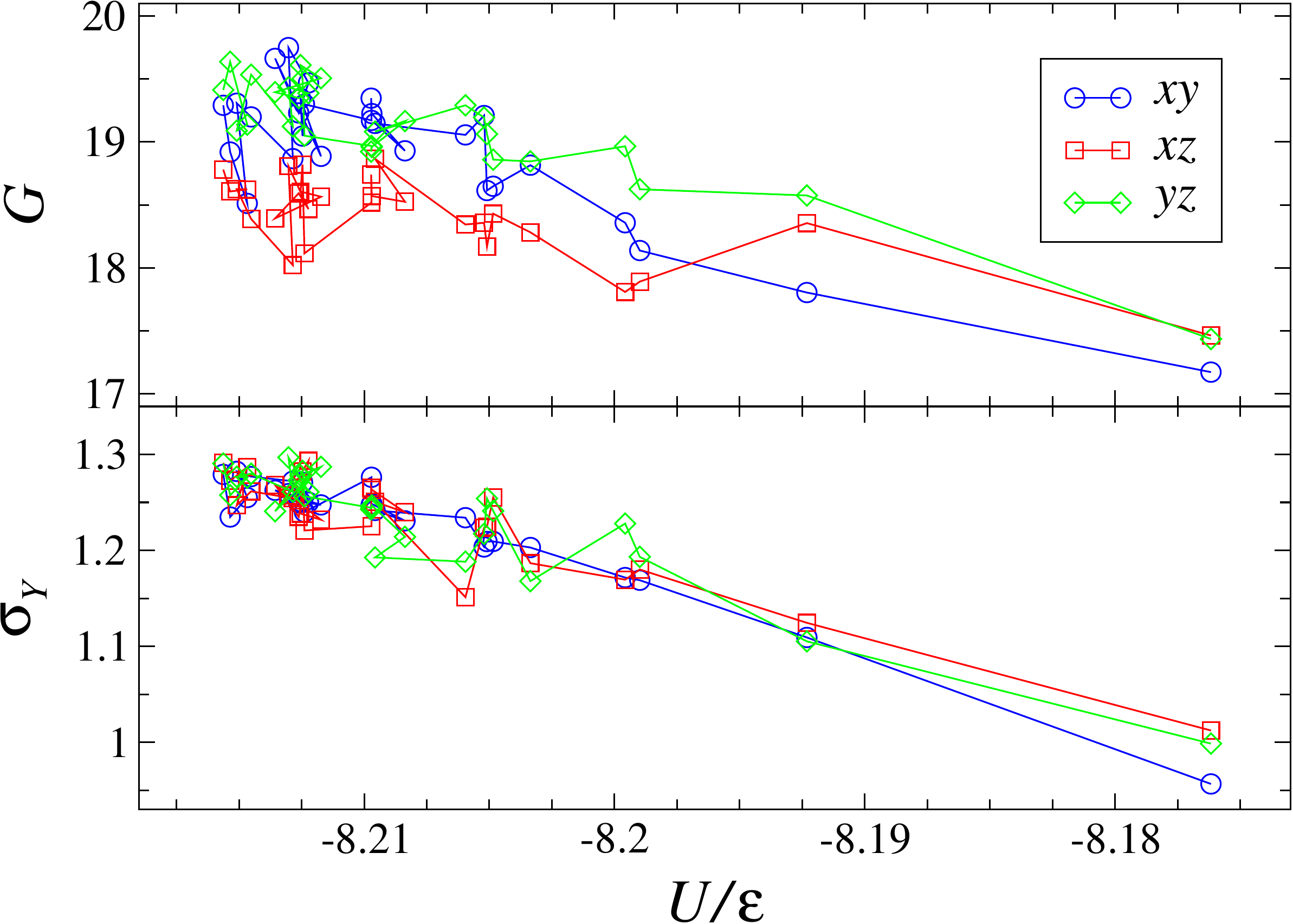}
\caption{(Color online)  The dependence of the shear modulus $G$ (in
units of $\varepsilon\sigma^{-3}$) and the yielding peak $\sigma_Y$
(in units of $\varepsilon\sigma^{-3}$) as a function of the
potential energy $U/\varepsilon$ (when strain is zero) at the
temperature $T_{LJ}=0.1\,\varepsilon/k_B$.  The glass is strained
along the $xy$ plane (blue circles), $xz$ plane (red squares), and
$yz$ plane (green diamonds). The strain rate is
$\dot{\gamma}=10^{-5}\,\tau^{-1}$. The same data as in
Figs.\,\ref{fig:poten_mech_ann} and \ref{fig:G_and_Y_T01}. }
\label{fig:G_and_Y_U_T01}
\end{figure}

We first evaluate the mechanical properties of steadily sheared
glasses and their dependence on the degree of annealing.  After the
amorphous system was brought from $T_{LJ}=0.30\,\varepsilon/k_B$ to
$0.10\,\varepsilon/k_B$ during $10^4\,\tau$, the shear modulus, $G$,
and the peak value of the stress overshoot, $\sigma_Y$, were
measured during startup continuous shear deformation with the strain
rate $\dot{\gamma}=10^{-5}\,\tau^{-1}$ at constant volume.  The
shear modulus was computed at strains $\gamma\leqslant0.01$ for each
shear orientation, \textit{i.e.}, along the $xz$, $xy$, and $yz$
planes. The results are presented in Fig.\,\ref{fig:G_and_Y_T01}. It
can be seen that both $G$ and $\sigma_Y$ gradually increase as a
function of the cycle number. This trend is consistent with the
decrease in potential energy at $T_{LJ}=0.10\,\varepsilon/k_B$
reported in Fig.\,\ref{fig:poten_mech_ann}. In other words, the
shear modulus and yield stress are greater in better annealed
glasses (see Fig.\,\ref{fig:G_and_Y_U_T01}). Note also that on
average $G_{xz}$ is smaller than $G_{xy}$ and $G_{yz}$.  The slight
shear modulus anisotropy arises due to a finite annealing time
($10^4\,\tau$) after the system was periodically deformed along the
$xz$ plane at $T_{LJ}=0.30\,\varepsilon/k_B$.

\vskip 0.05in

\begin{figure}[t]
\includegraphics[width=12.0cm,angle=0]{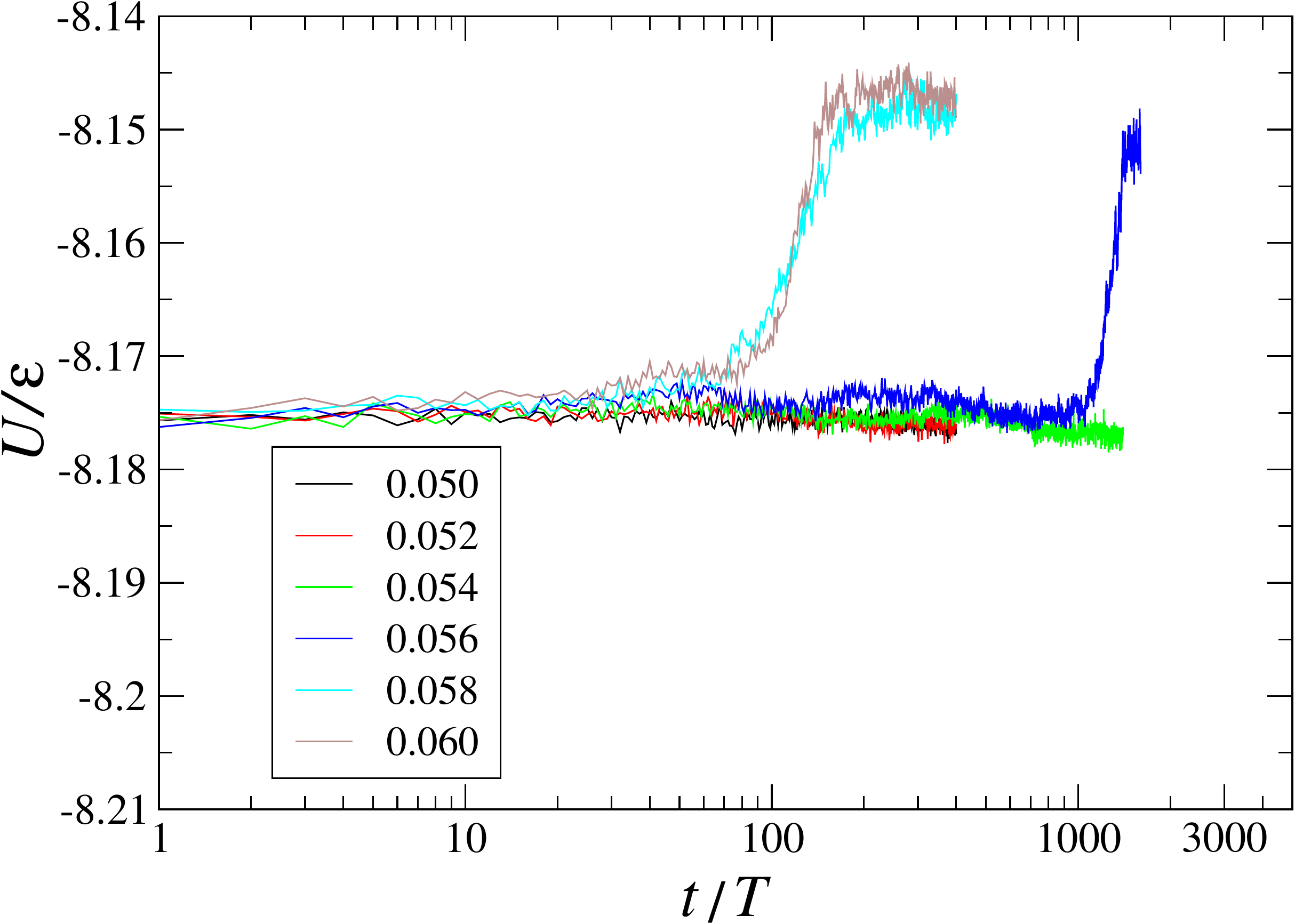}
\caption{(Color online) The time dependence of the potential energy
during periodic loading at $T_{LJ}=0.1\,\varepsilon/k_B$ for the
indicated values of the strain amplitude. The glass was initially
prepared by cooling with the rate $10^{-5}\varepsilon/k_{B}\tau$
from $T_{LJ}=1.0\,\varepsilon/k_B$ to $0.1\,\varepsilon/k_B$
(without mechanical annealing at $T_{LJ}=0.3\,\varepsilon/k_B$). The
period of shear deformation is $T=5000\,\tau$.}
\label{fig:poten_T01_just_prep_amp050_060}
\end{figure}
\begin{figure}[t]
\includegraphics[width=12.0cm,angle=0]{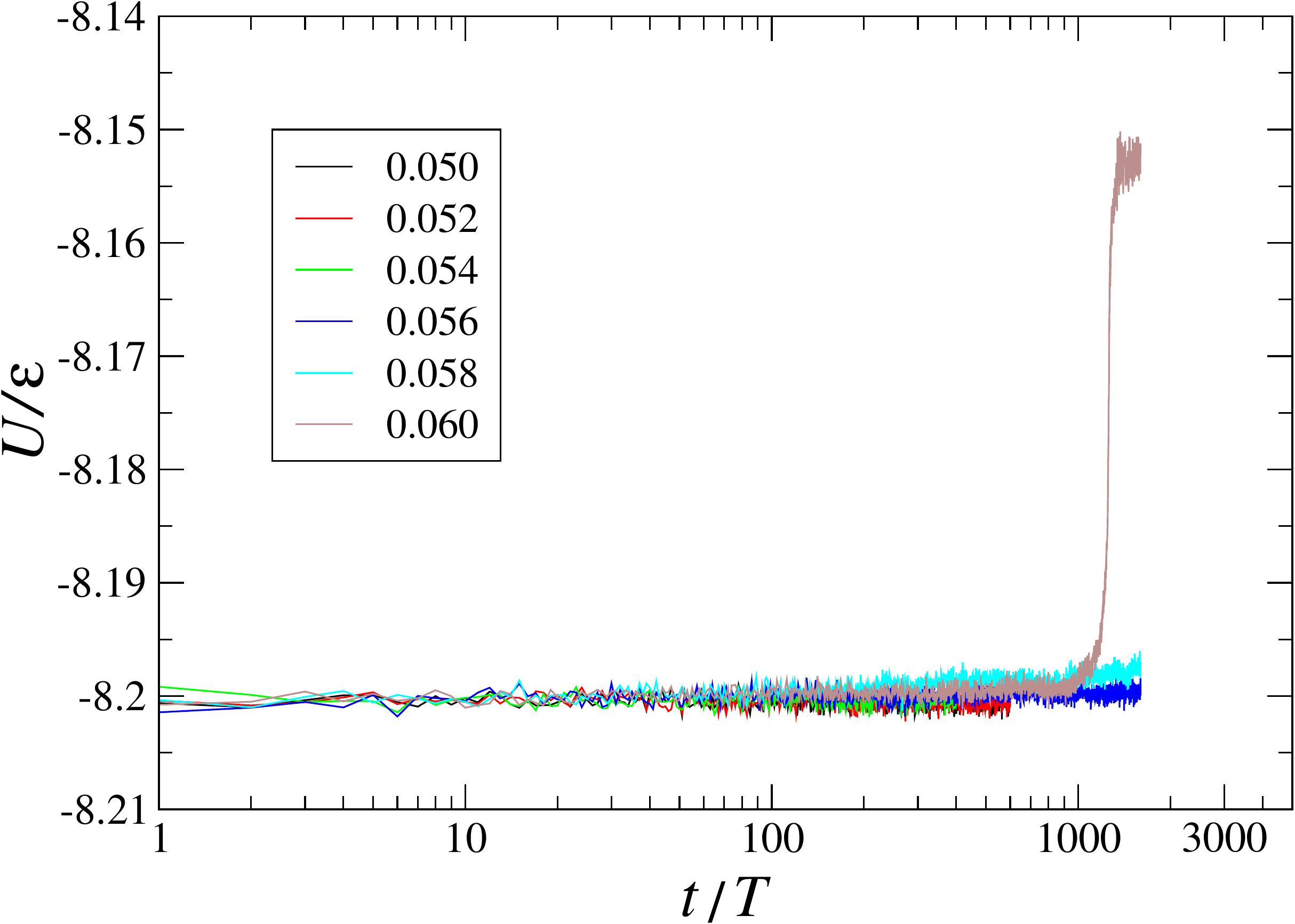}
\caption{(Color online) The variation of the potential energy minima
(at zero strain) during oscillatory shear at
$T_{LJ}=0.1\,\varepsilon/k_B$.  The values of the strain amplitude
are listed in the inset. The glass was first loaded during 300
cycles with $\gamma_0=0.035$ at $T_{LJ}=0.3\,\varepsilon/k_B$ and
then brought to the temperature $T_{LJ}=0.1\,\varepsilon/k_B$ (see
text for details). }
\label{fig:poten_T01_300cyc_amp050_060}
\end{figure}

The dynamic response of the binary glass to periodic shear
deformation with strain amplitudes in the range
$0.050\leqslant\gamma_0 \leqslant 0.060$ was first probed at the
temperature $T_{LJ}=0.10\,\varepsilon/k_B$. The variation of the
potential energy for thermally and mechanically annealed glasses are
shown in Fig.\,\ref{fig:poten_T01_just_prep_amp050_060} and
Fig.\,\ref{fig:poten_T01_300cyc_amp050_060}, respectively. In the
former case, the glass was cooled from $T_{LJ}=1.0\,\varepsilon/k_B$
to $0.1\,\varepsilon/k_B$ with the rate
$10^{-5}\varepsilon/k_{B}\tau$, while in the latter case, the glass
was first periodically strained during 300 cycles with
$\gamma_0=0.035$ at $T_{LJ}=0.3\,\varepsilon/k_B$ and only then
brought to $T_{LJ}=0.1\,\varepsilon/k_B$ during $10^4\,\tau$.  To
facilitate comparison, both vertical and horizontal scales in
Figs.\,\ref{fig:poten_T01_just_prep_amp050_060} and
\ref{fig:poten_T01_300cyc_amp050_060} are kept the same.  It can be
clearly observed that the potential energy (the plateau level before
yielding) of the mechanically annealed glass is reduced (by about
$\Delta U \approx 0.025\,\varepsilon$), and the critical strain
amplitude increases. More specifically, the critical strain
amplitude of the yielding transition is in the range $0.054 <
\gamma_0 < 0.056$ in Fig.\,\ref{fig:poten_T01_just_prep_amp050_060},
whereas the critical value becomes $0.058 < \gamma_0 < 0.060$ in
Fig.\,\ref{fig:poten_T01_300cyc_amp050_060}. Note that the upper
bounds for the critical strain amplitude were determined from the
sharp increase of the potential energy curves due to the formation
of a shear band. It should be emphasized that these conclusions hold
for periodic loading during 1600 cycles. In principle, the critical
value of the strain amplitude might be further reduced in both cases
if samples were deformed over additional cycles.

\vskip 0.05in


It should be mentioned that the characteristic increase in potential
energy above the critical strain amplitude, which is associated with
the formation of a system-spanning shear band, was repeatedly
reported in the previous MD
studies~\cite{Sastry17,Priezjev17,Priezjev18a,Sastry19band,Priez19ba,
Priez20alt,KawBer19,BhaSastry19}. In particular, it was demonstrated
that although the atomic density within a shear band is slightly
reduced and the potential energy locally increases, the glass
outside the shear band continues annealing upon subsequent periodic
deformation with an effectively lower strain
amplitude~\cite{Sastry19band}.   Generally, the formation of a shear
band at finite temperatures is delayed for cyclic loading above the
yielding strain amplitude for glasses prepared with slower cooling
rates~\cite{Priez19ba}.  More recently, it was also shown that
alternating shear orientation promotes the formation of shear bands
in amorphous alloys subjected to periodic
deformation~\cite{Priez20alt}.

\vskip 0.05in

\begin{figure}[t]
\includegraphics[width=12.0cm,angle=0]{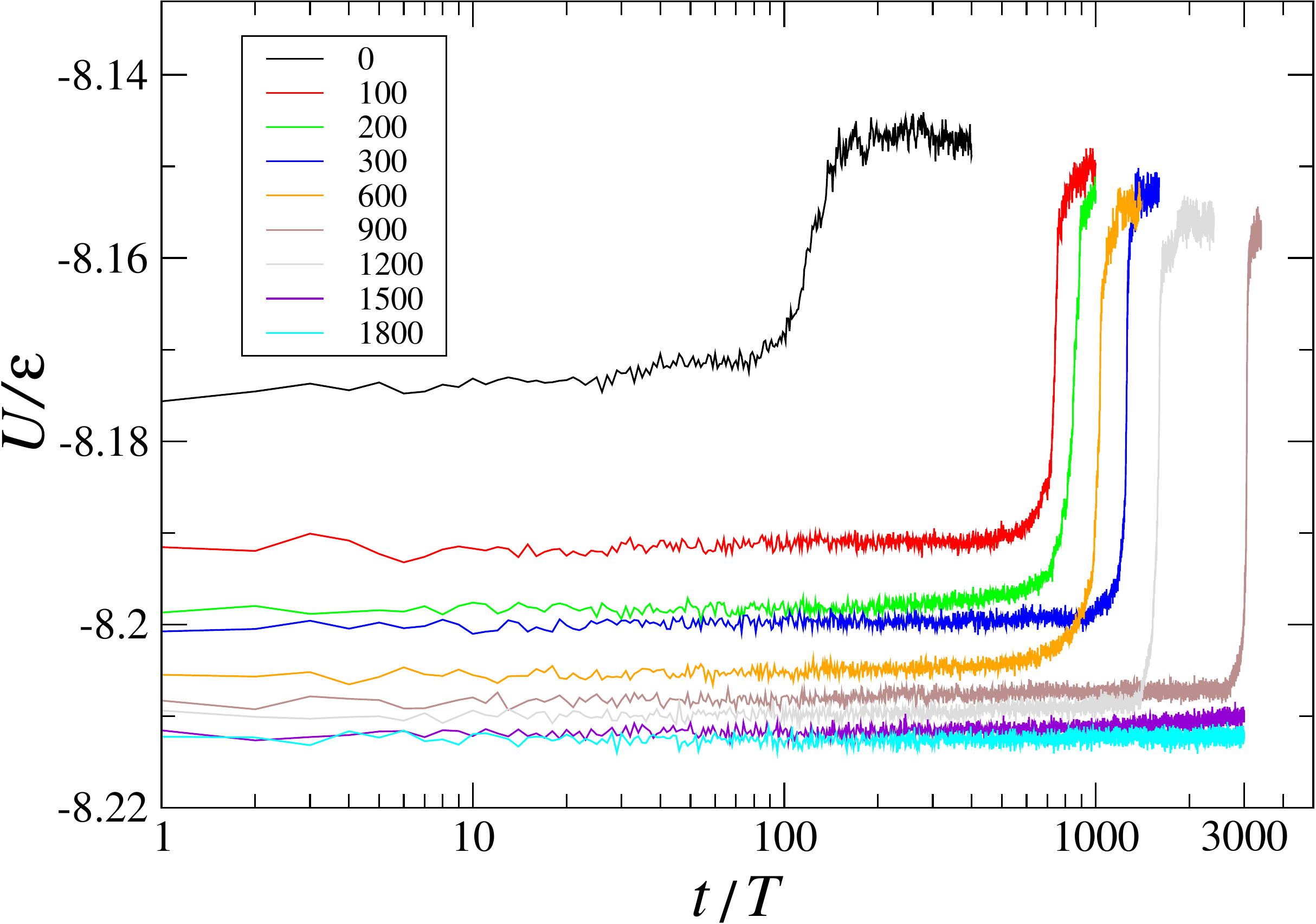}
\caption{(Color online) The potential energy versus cycle number for
the strain amplitude $\gamma_0=0.060$ and temperature
$T_{LJ}=0.1\,\varepsilon/k_B$. The number of cycles used for
mechanical annealing at $T_{LJ}=0.3\,\varepsilon/k_B$ and
$\gamma_0=0.035$ are listed in the legend. }
\label{fig:poten_T01_amp060}
\end{figure}

We next report the potential energy as a function of the number of
cycles at the strain amplitude $\gamma_0=0.060$ and temperature
$T_{LJ}=0.1\,\varepsilon/k_B$ in Fig.\,\ref{fig:poten_T01_amp060}.
Here, different curves denote the potential energy at the end of
each cycle for glasses that were periodically deformed at
$T_{LJ}=0.30\,\varepsilon/k_B$ for the specified number of cycles
(listed in the inset). Thus, the data for zero and 300 cycles are
same as in Figs.\,\ref{fig:poten_T01_just_prep_amp050_060} and
\ref{fig:poten_T01_300cyc_amp050_060}.  The results presented in
Fig.\,\ref{fig:poten_T01_amp060} demonstrate that the yielding
transition is delayed for glasses that were mechanically annealed to
lower energy levels. Note, however, that the number of cycles before
the transition is not always larger for lower energy states. For
example, the glass initially at $U\approx-8.20\,\varepsilon$ yields
after about 1250 cycles (the blue curve), while for
$U\approx-8.205\,\varepsilon$, the transition occurs after about
1000 cycles (the orange curve). Furthermore, the system dynamics
remains nearly reversible during 3000 cycles at the two lowest
energy states, leaving the possibility of yielding under continued
loading.

\vskip 0.05in

\begin{figure}[t]
\includegraphics[width=12.0cm,angle=0]{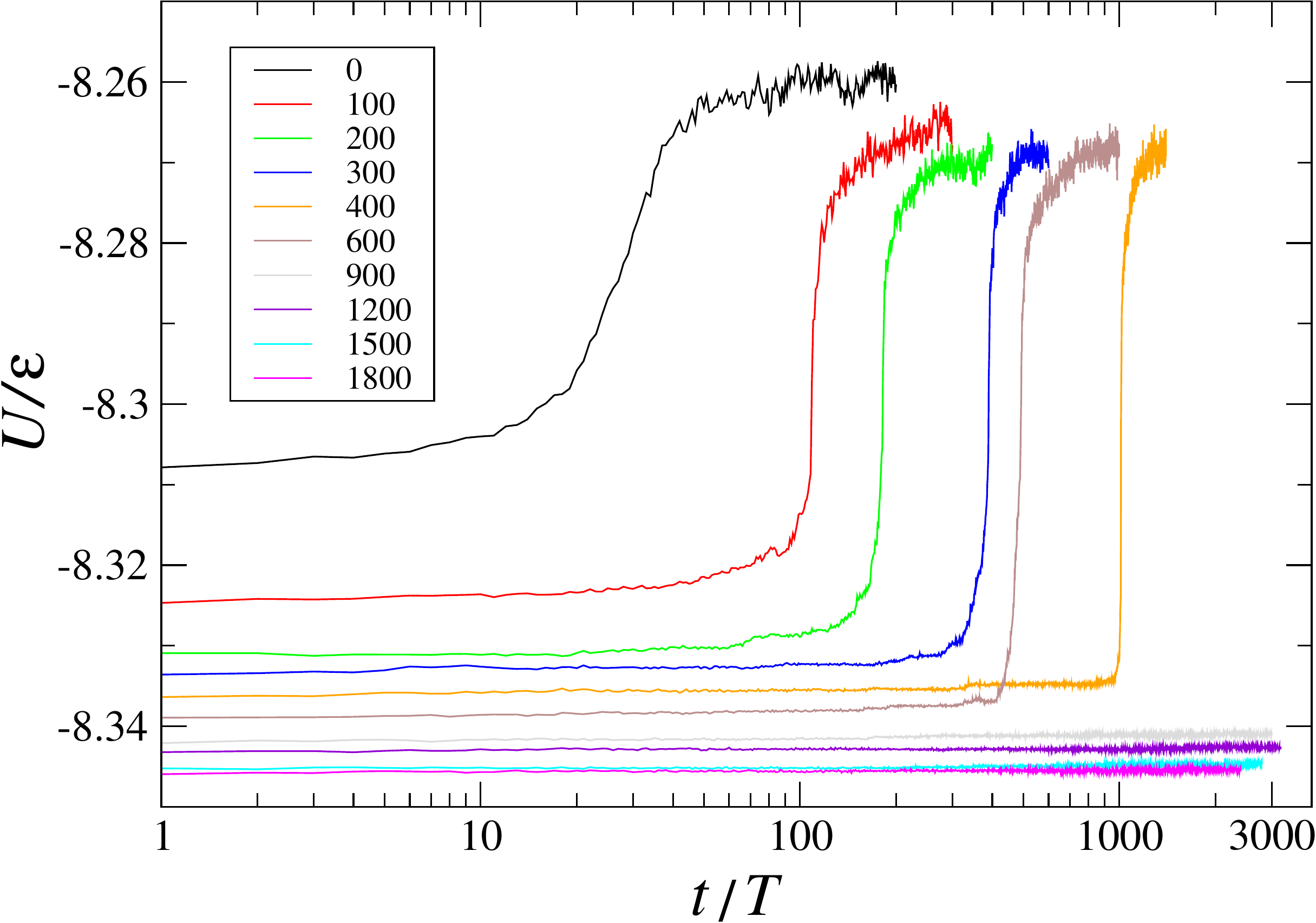}
\caption{(Color online) The dependence of the potential energy as a
function of the cycle number during periodic deformation with the
strain amplitude $\gamma_0=0.080$ at the temperature
$T_{LJ}=0.01\,\varepsilon/k_B$. The legend shows the number of
`annealing' cycles at $T_{LJ}=0.3\,\varepsilon/k_B$ (see text for
details). }
\label{fig:poten_T001_amp080}
\end{figure}

Similar results are also observed in
Fig.\,\ref{fig:poten_T001_amp080} for cyclic loading with the strain
amplitude $\gamma_0=0.080$ at the lower temperature
$T_{LJ}=0.01\,\varepsilon/k_B$. It was previously demonstrated that
the strain amplitude $\gamma_0=0.075$ is above the critical
amplitude for periodic shear deformation of the KA mixture at the
density $\rho=1.2\,\sigma^{-3}$ and cooling rates from $10^{-2}$ to
$10^{-5}\varepsilon/k_{B}\tau$~\cite{Priez19ba}. For reference, the
lowest potential energy in the KA glass cooled at
$10^{-2}\varepsilon/k_{B}\tau$ and then periodically deformed during
thousands of shear cycles at $T_{LJ}=0.01\,\varepsilon/k_B$ is
$U\approx -8.286\,\varepsilon$~\cite{Priez20alt}. As shown in
Fig.\,\ref{fig:poten_T001_amp080}, the number of cycles until
yielding increases roughly tenfold for lower energy glasses, whereas
at $U\lesssim-8.34\,\varepsilon$ the deformation remains reversible
during $\approx\!3000$ cycles.  Note that due to reduced thermal
fluctuations at $T_{LJ}=0.01\,\varepsilon/k_B$, the appearance of
discrete steps in the potential energy before the yielding
transition becomes more evident. As discussed below, the slight
increase in the potential energy is associated with nucleation of
small clusters of atoms with irreversible trajectories.

\vskip 0.05in


Further insights into the microscopic dynamics of atoms during
cyclic deformation can be gained by inspecting the so-called
nonaffine displacements. Briefly, the deviation from the affine
deformation for a group of atoms can be quantified via the nonaffine
measure $D^2(t, \Delta t)$, which is defined via the transformation
matrix $\mathbf{J}_i$ as follows:
\begin{equation}
D^2(t, \Delta t)=\frac{1}{N_i}\sum_{j=1}^{N_i}\Big\{
\mathbf{r}_{j}(t+\Delta t)-\mathbf{r}_{i}(t+\Delta t)-\mathbf{J}_i
\big[ \mathbf{r}_{j}(t) - \mathbf{r}_{i}(t)    \big] \Big\}^2,
\label{Eq:D2min}
\end{equation}
where $\Delta t$ is the time interval between successive positions
of $N_i$ atoms, and the sum is taken over nearest neighbors of the
$i$-th atom~\cite{Falk98}.  The spatiotemporal analysis on nonaffine
displacements of atoms was recently carried out to describe
irreversible dynamics in binary glasses under various loading
conditions, namely,
periodic~\cite{Priezjev16,Priezjev16a,Priezjev17,Priezjev18,Priezjev18a,
PriezSHALT19,Priez19ba,Priez20alt} and startup
continuous~\cite{Horbach16,Pastewka19,Priez20tfic,Priez19star,Ozawa20}
shear deformation, tension-compression
loading~\cite{NVP18strload,Jana20}, elastostatic
loading~\cite{PriezELAST19}, as well as thermal
cycling~\cite{Priez19one,Priez19tcyc,Priez19T2000,Priez19T5000}.

\vskip 0.05in

\begin{figure}[t]
\includegraphics[width=12.0cm,angle=0]{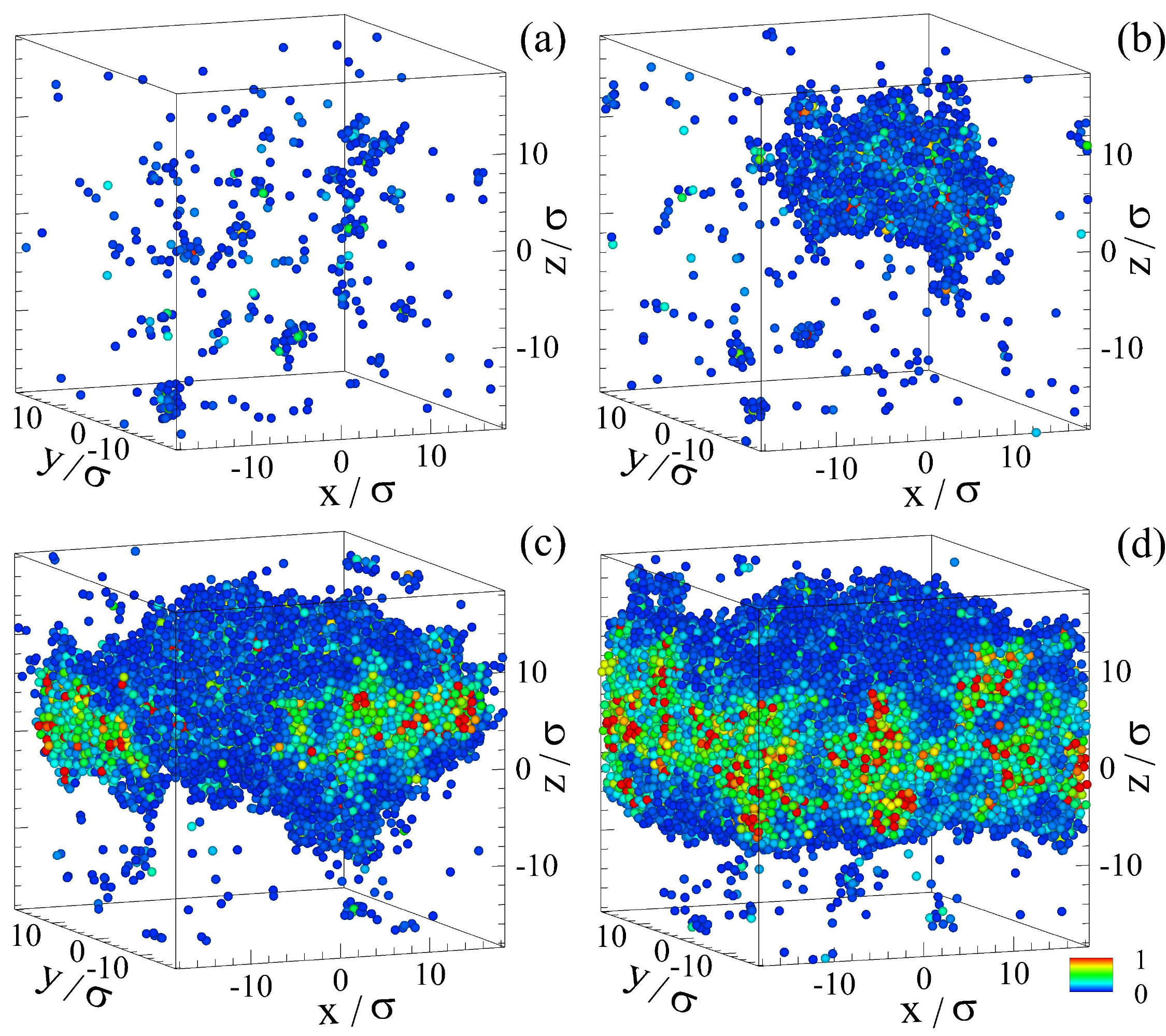}
\caption{(Color online) The selected snapshots of atomic
configurations during periodic shear with the strain amplitude
$\gamma_0=0.060$ at the temperature $T_{LJ}=0.1\,\varepsilon/k_B$.
The nonaffine measure is (a) $D^2(1000\,T, T)>0.04\,\sigma^2$, (b)
$D^2(1500\,T, T)>0.04\,\sigma^2$, (c) $D^2(1600\,T,
T)>0.04\,\sigma^2$, and (d) $D^2(2000\,T, T)>0.04\,\sigma^2$. The
legend indicates the magnitude of $D^2$. The oscillation period is
$T=5000\,\tau$. Atoms are not drawn to scale.  }
\label{fig:snapshots_kaa_1200cyc_T030_01_10000_amp060_D2}
\end{figure}

The sequence of snapshots of atomic configurations are presented in
Fig.\,\ref{fig:snapshots_kaa_1200cyc_T030_01_10000_amp060_D2}\,(a--d).
The data are taken during periodic deformation with the strain
amplitude $\gamma_0=0.060$ at $T_{LJ}=0.1\,\varepsilon/k_B$ (denoted
by the grey curve in Fig.\,\ref{fig:poten_T01_amp060}). The
corresponding energy levels along the grey curve are (a)
$U\approx-8.209\,\varepsilon$ at $t=1000\,T$, (b)
$-8.203\,\varepsilon$ at $1500\,T$, (c) $-8.179\,\varepsilon$ at
$1600\,T$, and (d) $-8.155\,\varepsilon$ at $2000\,T$. The parameter
$D^2(t,\Delta t = T)$ is computed for two consecutive configurations
at zero strain. For visualization of irreversible domains, only
atoms with relatively large nonaffine displacements,
$D^2(t,T)>0.04\,\sigma^2$, are displayed in
Fig.\,\ref{fig:snapshots_kaa_1200cyc_T030_01_10000_amp060_D2}. Note
that the typical cage size in the KA mixture at
$\rho=1.2\,\sigma^{-3}$ is about $0.1\,\sigma$, and, therefore,
empty regions in
Fig.\,\ref{fig:snapshots_kaa_1200cyc_T030_01_10000_amp060_D2}
correspond to spatial domains with nearly reversible dynamics.

\vskip 0.05in


It can be readily observed in
Fig.\,\ref{fig:snapshots_kaa_1200cyc_T030_01_10000_amp060_D2}\,(a)
that reversible deformation involves only a small number of isolated
atoms (mostly $B$ type) with large nonaffine displacements. In this
case, the reversible deformation of the well-annealed glass
continues for about 1300 cycles at the nearly constant energy level
$U\approx-8.209\,\varepsilon$.  In turn, the slight increase in
potential energy is associated with the nucleation of a cluster of
atoms with irreversible displacements, as shown in
Fig.\,\ref{fig:snapshots_kaa_1200cyc_T030_01_10000_amp060_D2}\,(b).
Furthermore, at about midway between two energy levels, the
formation of a shear band proceeds along the $xy$ plane via periodic
boundary conditions, as illustrated in
Fig.\,\ref{fig:snapshots_kaa_1200cyc_T030_01_10000_amp060_D2}\,(c).
Notice that the nonaffine measure varies gradually across the
partially formed shear band.  When the shear band is fully developed
in
Fig.\,\ref{fig:snapshots_kaa_1200cyc_T030_01_10000_amp060_D2}\,(d),
the potential energy levels out at the plateau,
$U\approx-8.155\,\varepsilon$, and the steady-state deformation
involves two well separated domains with diffusive and reversible
dynamics.

\vskip 0.05in

\begin{figure}[t]
\includegraphics[width=12.0cm,angle=0]{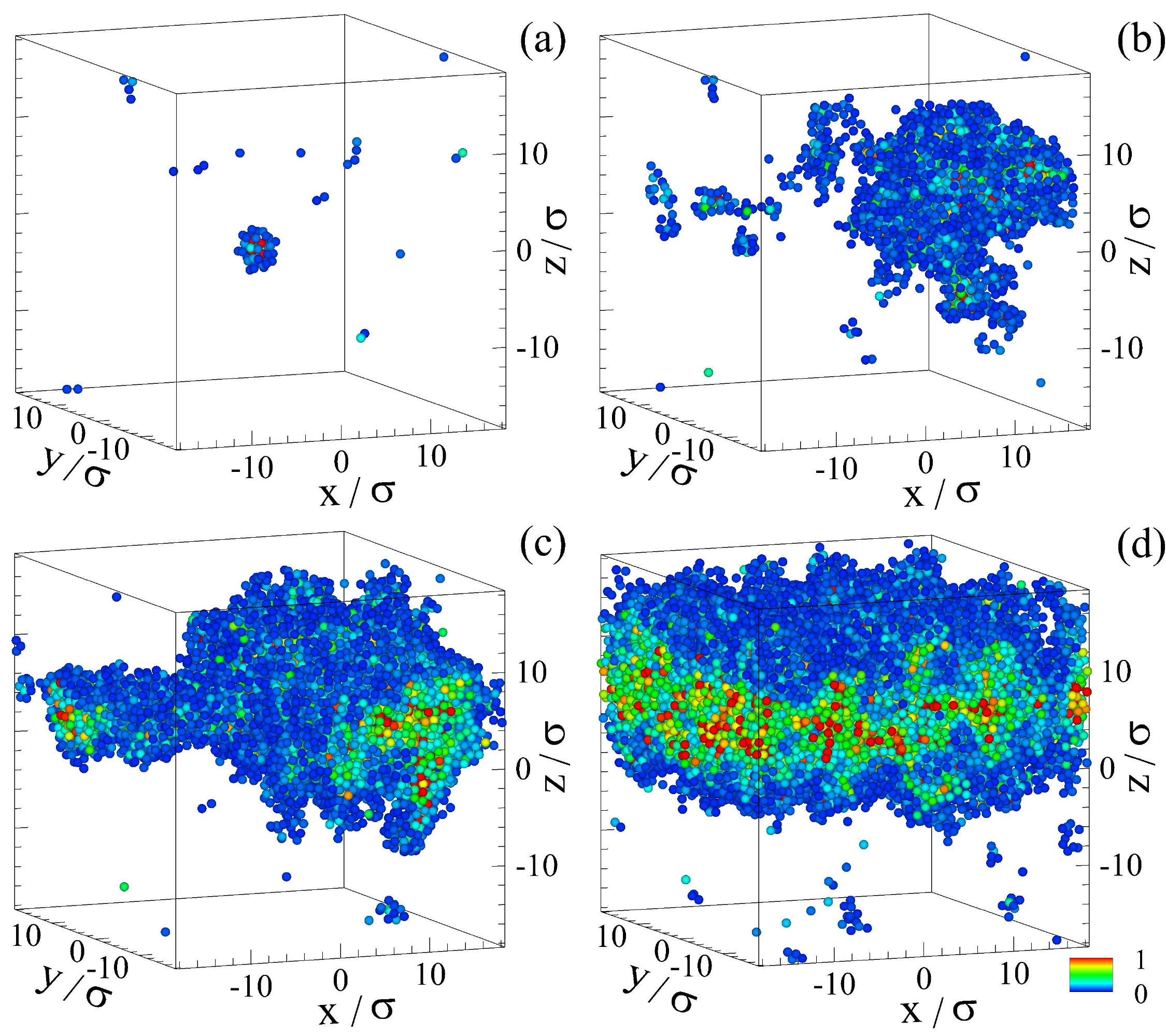}
\caption{(Color online) The positions of atoms at zero strain during
periodic loading with the strain amplitude $\gamma_0=0.080$ at the
temperature $T_{LJ}=0.01\,\varepsilon/k_B$. The nonaffine quantity
is (a) $D^2(300\,T, T)>0.04\,\sigma^2$, (b) $D^2(460\,T,
T)>0.04\,\sigma^2$, (c) $D^2(487\,T, T)>0.04\,\sigma^2$, and (d)
$D^2(800\,T, T)>0.04\,\sigma^2$. The magnitude of $D^2$ is defined
by the colorcode in the legend. }
\label{fig:snapshots_ka_600cyc_T030_001_10000_amp080_D2}
\end{figure}

Qualitatively similar results were observed during cyclic loading
with the strain amplitude $\gamma_0=0.080$ at the lower temperature
$T_{LJ}=0.01\,\varepsilon/k_B$. The selected snapshots in
Fig.\,\ref{fig:snapshots_ka_600cyc_T030_001_10000_amp080_D2}
illustrate the process of shear band formation in the well-annealed
glass described by the brown curve in
Fig.\,\ref{fig:poten_T001_amp080}. The potential energy values at
zero strain are (a) $U\approx-8.337\,\varepsilon$ at $t=300\,T$, (b)
$-8.328\,\varepsilon$ at $460\,T$, (c) $-8.310\,\varepsilon$ at
$487\,T$, and (d) $-8.268\,\varepsilon$ at $800\,T$. The first
indication of the approaching transition is reflected in the slight
increase of the potential energy after about 150 cycles (the brown
curve in Fig.\,\ref{fig:poten_T001_amp080}), which corresponds to
the formation of the cluster of about 30 atoms shown in
Fig.\,\ref{fig:snapshots_ka_600cyc_T030_001_10000_amp080_D2}\,(a).
When $U$ increases by about 15\% of the threshold energy at the
yielding transition, the cluster increases significantly but remains
compact, as displayed in
Fig.\,\ref{fig:snapshots_ka_600cyc_T030_001_10000_amp080_D2}\,(b).
Upon further loading, a tube-like structure is formed along the $y$
axis and then a narrow arm gets extended along the $x$ direction
[\,see
Fig.\,\ref{fig:snapshots_ka_600cyc_T030_001_10000_amp080_D2}\,(c)\,].
Finally, after the transition, periodically driven glass contains
the shear band along the $xy$ plane with the thickness of about half
the box size, as shown in
Fig.\,\ref{fig:snapshots_ka_600cyc_T030_001_10000_amp080_D2}\,(d).
Hence, the nucleation of small clusters of atoms with irreversible
trajectories and the corresponding discrete steps in the potential
energy curves demonstrate that the location of a shear band across a
system can be predicted at least several cycles before the yielding
transition.

\section{Conclusions}

In summary, molecular dynamics simulations were performed to
investigate the influence of glass stability, temperature, and
strain amplitude on reversibility and yielding transition in
cyclically loaded amorphous materials. The model glass consists of a
mixture of two types of atoms with strongly non-additive
interactions, which forms a disordered solid when gradually cooled
below the glass transition temperature. A wide range of potential
energy states were accessed via small-amplitude periodic deformation
during thousands of shear cycles. It was shown that both the shear
modulus and peak value of the stress overshoot increase in more
stable glasses.  Furthermore, we found that the critical strain
amplitude increases in glasses mechanically annealed to lower
potential energies when periodically deformed for about one thousand
shear cycles at a finite temperature.  At the same time, if the
strain amplitude is fixed, the yielding transition is delayed in
highly annealed glasses deformed over thousands of cycles. The
spatial analysis of nonaffine displacements elucidates the process
of strain localization, which includes a nucleation of a small
cluster of atoms with irreversible trajectories, followed by
propagation of a shear band along a plane.

\section*{Acknowledgments}

Financial support from the National Science Foundation (CNS-1531923)
is gratefully acknowledged.  The article was prepared within the
framework of the HSE University Basic Research Program and funded in
part by the Russian Academic Excellence Project `5-100'. The
numerical simulations were carried out at Wright State University's
Computing Facility and the Ohio Supercomputer Center. The molecular
dynamics simulations were carried out using the efficient LAMMPS
code developed at Sandia National Laboratories~\cite{Lammps}.



%

%

%

%

%

%

%

%

%

\bibliographystyle{prsty}

\end{document}